\documentclass[a4paper,11pt]{article}
\pdfoutput=1 

\usepackage{jinstpub} 
\usepackage{enumerate}

\usepackage{lineno}

\title{PMT gain calibration and monitoring based on highly compressed hit information in KM3NeT}

\author[a,b]{Bouke Jung,\note{Corresponding author.}}
\author[a,c]{Maarten de Jong}
\author[d]{Paolo Fermani}

\affiliation[a]{Dutch National Institute for Subatomic Physics (Nikhef),\\
  Science Park 105, Amsterdam, The Netherlands}

\affiliation[b]{University of Amsterdam, Institute of Physics,\\
  Science Park 904, Amsterdam, The Netherlands}

\affiliation[c]{Leiden University, Leiden Institute of Physics,\\
  Niels Bohrweg 2, Leiden, The Netherlands}

\affiliation[b]{Sapienza Universit\`{a} di Roma, Dipartimento di Fisica,\\
Piazzale Aldo Moro 5, Rome, Italy}

\emailAdd{bjung@nikhef.nl}

\abstract{The cubic-kilometre neutrino telescope, which consists of large-scale 3D-arrays of photomultiplier tubes (PMTs) currently under construction on the Mediterranean seabed, relies on accurate calibration procedures in order to answer its science goals. These proceedings present the gain calibration method used in KM3NeT, which is based on highly compressed PMT hit information. In particular, it is shown that the PMT gains can be tuned to within 2\% of the nominal value, based on the measured single photoelectron time-over-threshold distribution of each PMT.}


\collaboration[c]{on behalf of the KM3NeT collaboration}

\proceeding{Very Large Volume Neutrino Telescope 2021 conference\\
  18-21 May, 2021\\
  Valencia / online}

\newcommand{\pe}{\,\text{p.e.}}
\newcommand{\ns}{\,\text{ns}}
\newcommand{\V}{\,\text{V}}

\newcommand\numberthis{\addtocounter{equation}{1}\tag{\theequation}}

\begin{document}
\maketitle
\flushbottom

\section{Introduction}
\label{section:Introduction}

The cubic-kilometre neutrino telescope (KM3NeT) is a deep-sea neutrino research infrastructure comprising two large-volume water Cherenkov detectors which consist of 3D-arrays of optical modules, each one containing 31 3-inch photomultiplier tubes (PMTs) \citep{Margiotta2014}. The first detector, called KM3NeT-ARCA, is optimised for cosmic neutrino sources \citep{Aiello2019}, whilst the second one, called KM3NeT-ORCA, is optimised for the determination of the neutrino mass ordering \citep{Aiello2021}. To reach the science goals of the experiment, accurate calibration of the PMTs is essential. The timing and quantum efficiency calibration of the PMTs has been described in previous papers \citep{Melis2017, Coniglione2019}. These proceedings present an overview of the PMT gain calibration procedure used in KM3NeT. In contrast to other gain calibration procedures \citep{Aartsen2020, Amaudruz2019, Abe2014}, the method does not depend on the output of analogue-to-digital-converters (ADCs), which are not built into the KM3NeT readout electronics \citep{Aiello2021_2}. Instead, it is based on the output of a time-to-digital-converter (TDC), which records the arrival time of a PMT signal and the duration over which the analogue pulse exceeds the voltage threshold set by the hardware discriminator, commonly referred to as the time-over-threshold. Combined with the PMT identifier, this allows all PMT signal information to be sent to shore in highly compressed 6-Byte data-formats (hits), which avoids the need for prior data filtering, in line with the all-data-to-shore principle which KM3NeT has adopted \citep{Pellegrino2016}.





\section{The PMT analogue pulse model}
\label{section:pulseShapeModel}

The gain calibration of the PMTs in KM3NeT is based on an analytical model, which describes the dependence of the time-over-threshold of a PMT hit, $\Delta T$, on the integrated charge $q$ of the corresponding analogue pulse. This model is based on the assumption that the analogue pulse follows the shape of a Gaussian with standard deviation $\sigma$ and transitions into an exponential tail that decays at a rate $\tau$. The time-over-threshold can be calculated as the difference between the first and last time during which the modeled pulse exceeds the voltage threshold, $V_{0}$, set by the discriminator. In this calculation, the saturation effect mentioned in Ref. \citep{Reubelt2019} is taken into account as an effective linear extension of the time-over-threshold as a function of the charge above $1 \pe$.\footnote{The second saturation for pulses with an integrated charge above roughly $30\pe$ is also modeled, but can be safely ignored for the purpose of PMT gain calibration, where only single photoelectron signals are used.} This gives rise to the following definition of the time-over-threshold as a function of the charge:

\begin{align*}
  \Delta T(q) 
  &= 
  \begin{cases}
    2 \sigma \sqrt{2 \ln\left(\frac{q}{q_0}\right)} , &q_0 < q \leq \frac{q_0}{C}, \\
    \tau \ln\left(\frac{q}{q_0 \cdot C}\right) + \sigma \sqrt{2 \ln\left(\frac{q}{q_0}\right)}, &\frac{q_0}{C} < q \leq q_{\mathrm{L}}, \\
    \tau \ln\left(\frac{q_{\mathrm{L}}}{q_0 \cdot C}\right) + \sigma \sqrt{2 \ln\left(\frac{q_{\mathrm{L}}}{q_0}\right)} + \beta (q-q_{\mathrm{L}}), &q > q_{\mathrm{L}}, \\
    0, &q \leq q_0,
  \end{cases} \numberthis
\label{eq:ToTwithoutSaturation}
\end{align*}

\noindent where $C = e^{-\frac{1}{2}\left(\frac{\sigma}{\tau}\right)^2}$ is the signal amplitude at the transition point between the exponential and Gaussian regimes of the pulse, $q_{0} \approx 0.24\pe$ is the threshold-equivalent charge, $\beta$ is the derivative of the time-over-threshold computed with respect to the charge in the linear regime and $q_{L} \approx 1.36\pe$ is the lower boundary of the linear regime, set by the condition $\frac{\partial (\Delta T)}{\partial q} \rvert_{q = q_{L}} = \beta$. Experimentally, the value of $\beta$ has been measured to be $7.0 \ns \pe^{-1}$ \citep{Schermer2017}. The full behaviour of the time-over-threshold as a function of the charge, given by equation \ref{eq:ToTwithoutSaturation}, is shown on the left-hand side of figure \ref{fig:ToT}.

\section{Time-over-threshold-based PMT gain calibration}
\label{section:ToTandGainFit}

To relate the PMT gain to the time-over-threshold, a definition of the single photoelectron charge distribution must be given. In KM3NeT, this distribution is modeled as a two-component Gaussian mixture, accounting for nominally amplified PMT pulses corresponding to a signal amplification of $3 \times 10^{6}$ \citep{Aiello2018} and so-called underamplified pulses, arising from diverse effects such as photons which pass through the photocathode and impinge on the first dynode, photoelectrons which skip certain dynode stages or photoelectrons which backscatter inelastically on the first dynode \citep{Carter1980, Aiello2010}:

\begin{align}
  f(q) &= \frac{1}{A} \left( p \cdot \mathcal{G}(q; \mu_{u},\sigma_{u}^2) + (1-p) \cdot \mathcal{G}(q; \mathrm{G},\Sigma^2) \right).
  \label{eq:SPEdistr}
\end{align}

\noindent In this equation, $A$ is a normalisation constant which accounts for the truncation of the distribution by the hardware threshold, $p$ corresponds to the occurrence probability of underamplified pulses, $\mathcal{G}(q; G,\Sigma^2)$ corresponds to the nominally amplified component and $\mathcal{G}(q; \mu_u,\sigma_u^2)$ corresponds to the underamplified contribution. The variance of the nominally amplified distribution $\Sigma$ is determined via the gain $G$ and gainspread $\sigma_{G}$ as $\Sigma = \sigma_G \sqrt{G}$. Conversely, the mean and standard deviation of the underamplified distribution are defined as $\mu_u = \sigma_G^2 \cdot G = \Sigma^2$ and $\sigma_u = \sigma_G \cdot \Sigma$. An analytical expression for the shape of the single photoelectron time-over-threshold distribution can be derived from the charge distribution defined in equation \ref{eq:SPEdistr} by means of the conversion model defined in equation \ref{eq:ToTwithoutSaturation}:

\begin{figure}[!ht]
  \centering
  \includegraphics[width=0.49\textwidth]{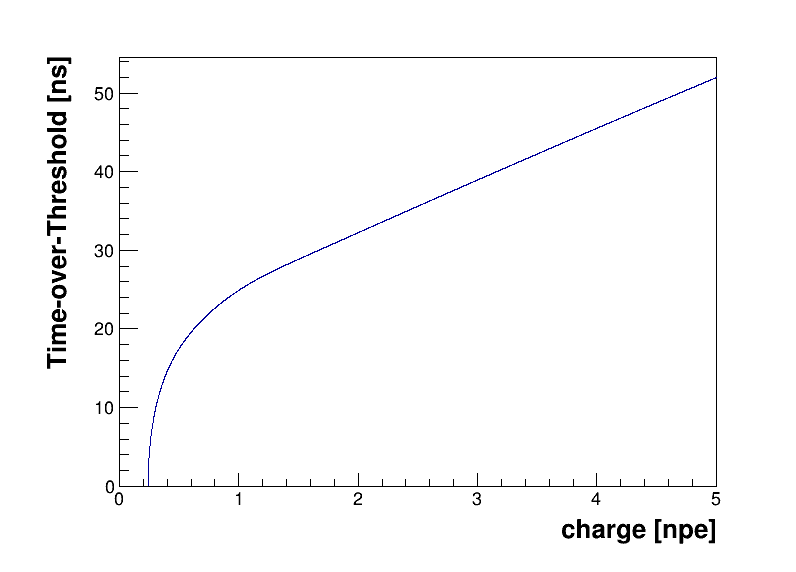}
  \includegraphics[width=0.49\textwidth]{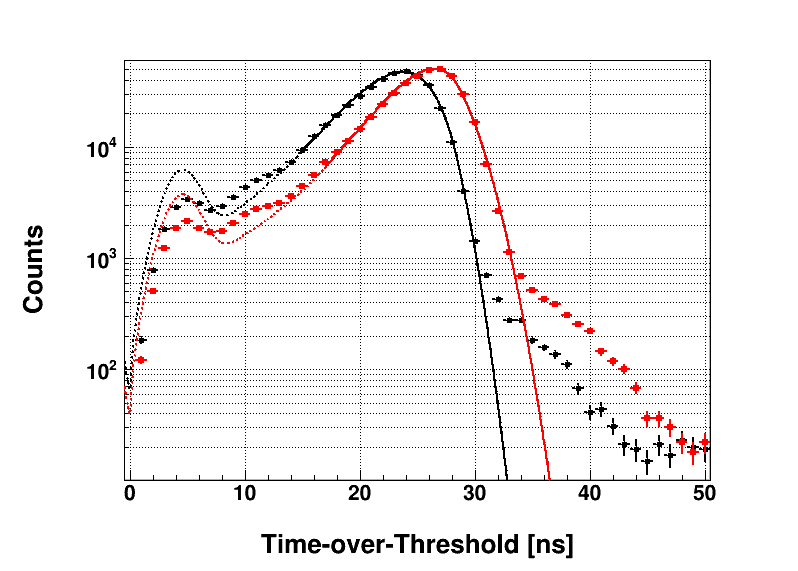}
  \caption{\small The time-over-threshold as function of the charge, as described by equation \ref{eq:ToTwithoutSaturation} (left) and two time-over-threshold distributions measured for a single PMT operating at the ORCA site under high voltages of $-1130 \V$ (black) and $-1180 \V$ (red) and fitted according to equation \ref{eq:ToTdistr} (right). For the right figure, solid lines indicate the fit region. Extrapolations for the full domain are indicated with dotted lines.}
  \label{fig:ToT}
\end{figure}

\begin{equation} 
  f(\Delta T) = W_{\mathrm{a}} \cdot f(q(\Delta T)) \cdot \frac{\partial q(\Delta T)}{\partial \Delta T}.
  \label{eq:ToTdistr}
\end{equation}

\noindent In this equation, $W_{\mathrm{a}} = \int_{q_0}^{\infty} f(q) dq$. By fitting multiple single photoelectron time-over-threshold distributions taken at different high-voltage settings as a function of the PMT gain, the high-voltage which gives rise to a nominal gain can be determined. Figure \ref{fig:ToT} shows two time-over-threshold distributions generated for the same PMT deployed at the ORCA-site under two different high-voltage settings. The data were obtained by filtering out signals from $^{40}$K-decays, which are dominated by single photoelectron signals. Subdominant contributions from signals generated by the release of multiple primary photoelectrons from the photocathode can be seen at time-over-threshold values greater than $30\pe$. The fits of the model to the data are indicated with solid lines. Also visible are extrapolations of the model over the full time-over-threshold domain. In these extrapolations, an additional Gaussian component can be distinguished which has recently been added to equation \ref{eq:ToTdistr} in an attempt to capture the small excess of pulses typically observed at very low time-over-threshold values, below $7\ns$. In order to avoid any biases, this additional component is currently not included in the fits during calibration. Ways to improve the modeling of the lower end of the time-over-threshold distributions are under investigation. 




\begin{figure}[!ht]
  \centering
  \includegraphics[width=0.49\textwidth]{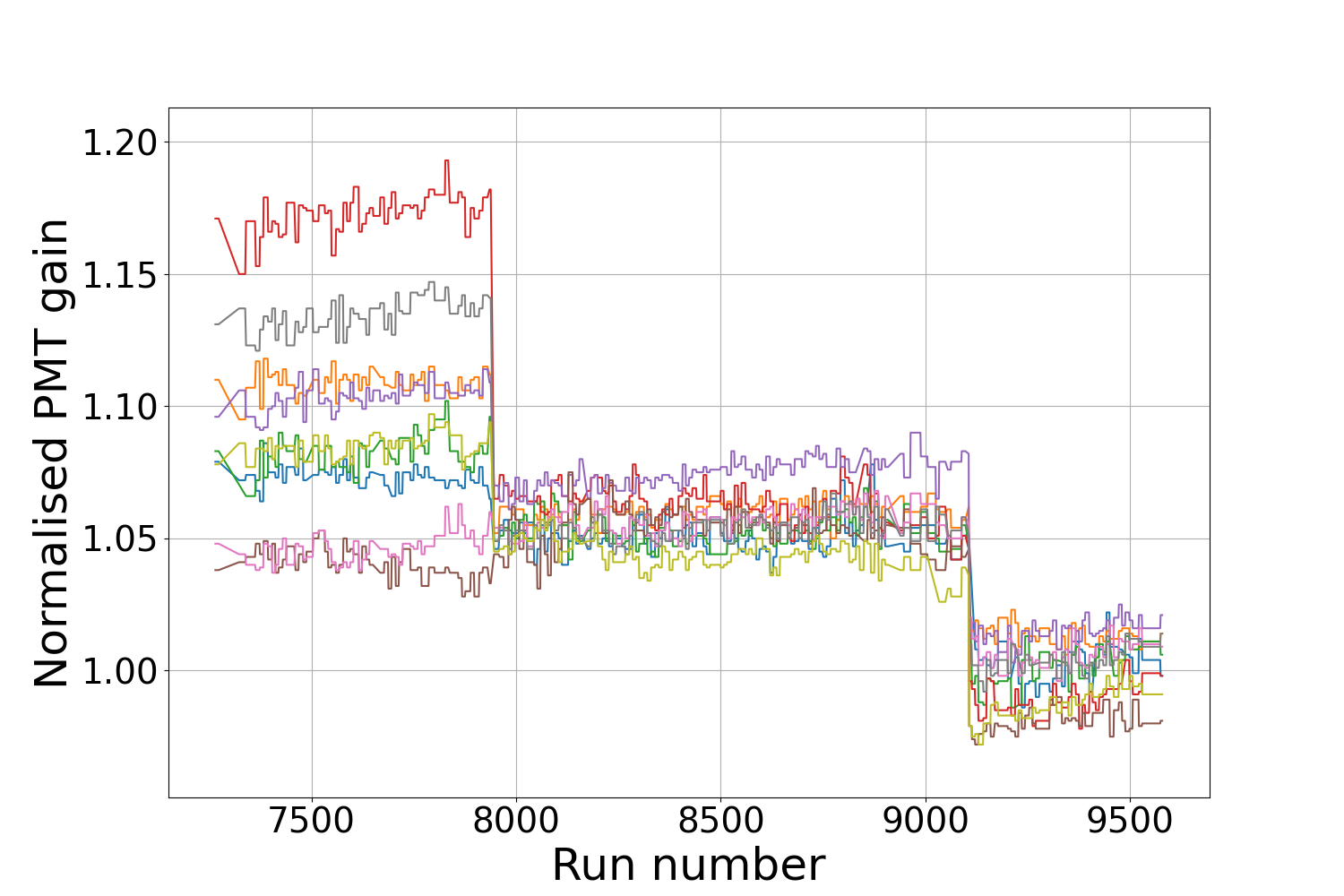}
  \includegraphics[width=0.49\textwidth]{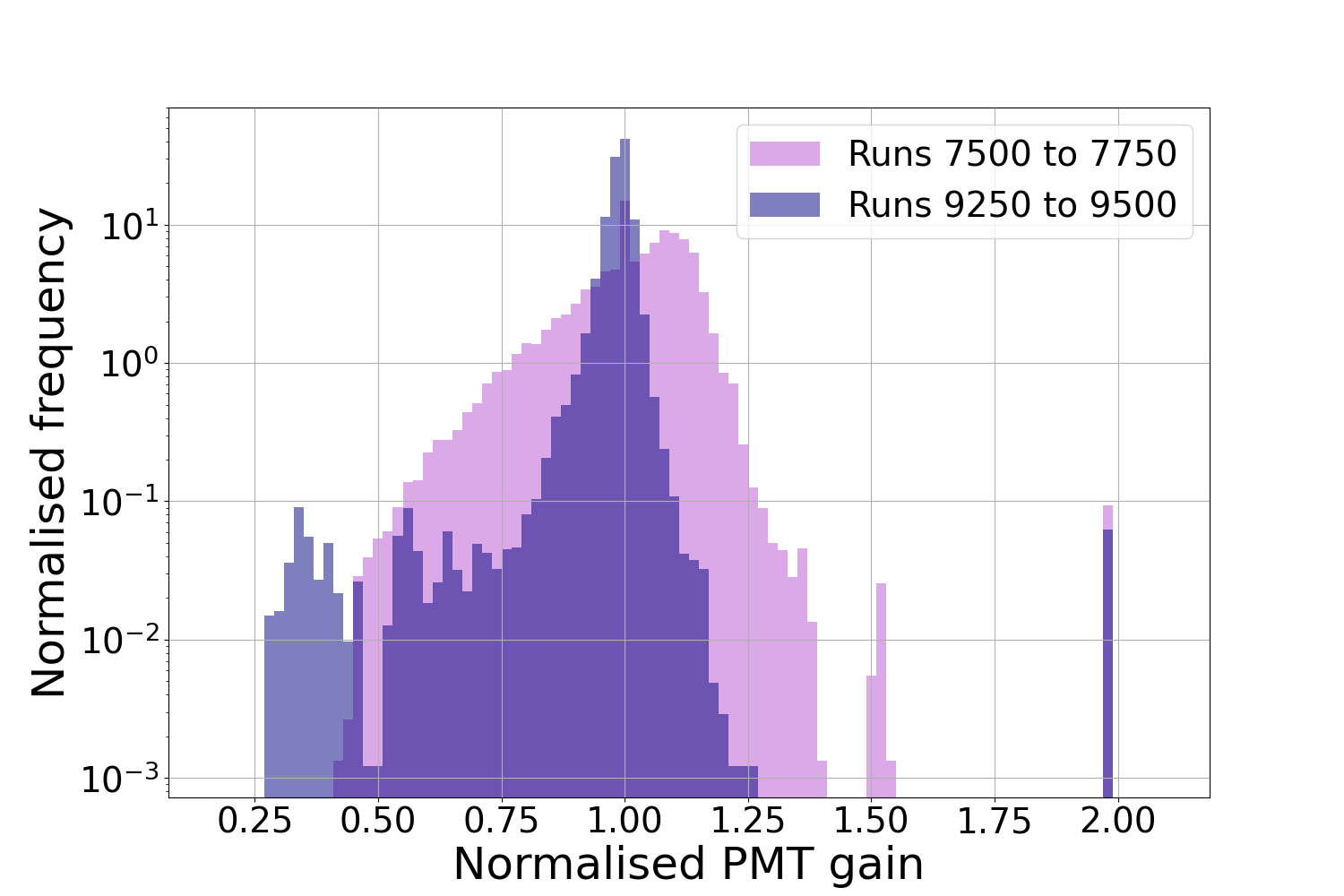}
  \caption{\small The PMT gains normalised with respect to the nominal amplification of $3 \times 10^{6}$ for a selection of 9 PMTs in a string deployed at the ORCA site, monitored between January 26th, 2020 and March 3rd, 2021 (left) and the distribution of PMT gains for two distinct data-taking periods, normalised over the number of runs and total number of operating PMTs (right).}
  \label{fig:gainMonitoring}
\end{figure}


\par \noindent The left-hand side of figure \ref{fig:gainMonitoring} shows the gain evolution for a selection of 9 PMTs deployed at the ORCA site, monitored over a period of roughly one year. Two HV-tuning campaigns are visible, which were initiated to equalise the PMT gains. These campaigns have significantly reduced the spread in the gain values, as can be seen on the right-hand side of figure \ref{fig:gainMonitoring}. A slight overestimation of the gain can be seen after the first application of the new gain calibration procedure, between runs 7950 and 9100. During this period, additional improvements in the software were made, which resulted in an equalisation of the normalised PMT gains to within 2\% of the nominal value of 1.0 after the subsequent calibration. The appearance of the small subpopulations below a normalised gain of 0.75 and close to the maximum of the gain fit range of 2.0 after the second tuning on the right-hand side of figure \ref{fig:gainMonitoring}, are caused by PMTs for which no valid high-voltage setting could be determined based on the current calibration procedure. The high-voltage for these PMTs were reset to the value recommended by the vendor, which are known to result in sub-optimal gains in the sea.


\section{Conclusions}
\label{section:conclusions}


These proceedings have outlined the method used for the calibration of PMT gains in KM3NeT, based on the time-over-threshold information. The model used to fit measured time-over-threshold distributions as a function of the gain has been explained. Using this model, the PMT gains in KM3NeT can be equalised to within 2\% of the nominal value.





\bibliographystyle{JHEP}\it
{\small \setlength{\bibsep}{1.0pt}
\bibliography{VLVnT_2021_bjung_PMTgainCalibrationAndMonitoring_proceedings}}






\end{document}